\documentclass[10pt,journal,compsoc]{IEEEtran}

\ifCLASSOPTIONcompsoc
  \usepackage[nocompress]{cite}
\else
  \usepackage{cite}
\fi

\ifCLASSINFOpdf
  \usepackage[pdftex]{graphicx}
\else
  \usepackage[dvips]{graphicx}
\fi

\ifCLASSOPTIONcompsoc
  \usepackage[caption=false,font=footnotesize,labelfont=sf,textfont=sf]{subfig}
\else
  \usepackage[caption=false,font=footnotesize]{subfig}
\fi

\ifCLASSOPTIONcaptionsoff
  \usepackage[nomarkers]{endfloat}
 \let\MYoriglatexcaption\caption
 \renewcommand{\caption}[2][\relax]{\MYoriglatexcaption[#2]{#2}}
\fi

\usepackage{amsmath}
\usepackage{amssymb}
\usepackage{empheq}

\usepackage{tikz}
\usepackage{pgfplots}
\usetikzlibrary{shapes,arrows,calc,patterns,decorations.pathmorphing,decorations.markings,circuits.ee.IEC}
\usepackage{circuitikz}
\usetikzlibrary{calc}

\usepackage[printonlyused,smaller]{acronym}

\acrodef{AAC}{Advanced Audio Coding}
\acrodef{ADC}{Analog Digital Converter}
\acrodef{AGU}{Address Generating Unit}
\acrodef{ALU}{Algorithmic Logical Unit}
\acrodef{AVM}{Astute Virtual Machine}
\acrodef{ASIC}{Application-Specific Integrated Circuit}
\acrodef{BEEBS}{Bristol Energy Efficiency Benchmark Suite}
\acrodef{BSIM}{Berkeley Short-channel IGFET Model}
\acrodef{BTBT}{band-to-band tunneling}
\acrodef{CISC}{Complex Instruction Set Computer}
\acrodef{CMOS}{complementary metal-oxide-semiconductor}
\acrodef{CFD}{computational fluid dynamics}
\acrodef{CPU}{Central Processing Unit}
\acrodef{DFG}{Data Flow Graph}
\acrodef{DDIO}{Data Direct I/O}
\acrodef{DLP}{Data Level Parallelism}
\acrodef{DMA}{Direct Memory Access}
\acrodef{DRAM}{Dynamic Random-Access Memory}
\acrodef{DSP}{Digital Signal Processor}
\acrodef{DTM}{Dynamic Thermal Management}
\acrodef{DVS}{Dynamic Voltage Scaling}
\acrodef{DVFS}{Dynamic Voltage and Frequency Scaling}
\acrodef{DPM}{Dynamic Power Management}
\acrodef{EDD}{Energy-Delay Diagram}
\acrodef{EEPROM}{Electrically Erasable Programmable Read-Only Memory}
\acrodef{EER}{Energy Efficiency Rating}
\acrodef{FFT}{Fast Fourier Transformation}
\acrodef{FPA}{Floating Point Adder}
\acrodef{FPU}{Floating Point Unit}
\acrodef{FPM}{Floating Point Multiplier}
\acrodef{FMA}{fused multiply?add}
\acrodef{FSM}{Finite State Machine}
\acrodef{GPU}{Graphics Processing Unit}
\acrodef{GPS}{Global Positioning System}
\acrodef{GSM}{Global System for Mobile Communications}
\acrodef{HC}{Hardware Counter}
\acrodef{HDL}{Hardware Description Language}
\acrodef{HPC}{High Performance Computing}
\acrodef{IC}{Integrated Circuit}
\acrodef{IR}{infrared}
\acrodef{IoT}{Internet of Things}
\acrodef{ILP}{integer linear programming}
\acrodef{I2C}{Inter-Integrated Circuit}
\acrodef{IQR}{interquartile range}
\acrodef{ITRS}{In\-ter\-na\-tion\-al Tech\-nolo\-gy Road\-map for Semi\-con\-duc\-tors}
\acrodef{ILP}{Instruction Level Parallelism}
\acrodef{I/O}{input/output}
\acrodef{ISA}{Instruction Set Architecture}
\acrodef{FIR}{finite impulse response}
\acrodef{JIT}{Just-In-Time}
\acrodef{JNI}{Java Native Interface}
\acrodef{LPDDR}{low power DRAM}
\acrodef{LCD}{liquid crystal display}
\acrodef{MAD}{median absolute deviation}
\acrodef{MAE}{maximum absolute error}
\acrodef{MIPJ}{millions-of-instructions-per-joule}
\acrodef{MOSFET}{metal-oxide semiconductor field-effect transistor}
\acrodef{MPSoC}{multiprocessor System-on-Chip}
\acrodef{MTTF}{Mean Time To Failure}
\acrodef{MIPS}{Microprocessor without Interlocked Pipeline Stages}
\acrodef{NIC}{Network Interface Card}
\acrodef{NIST}{National Institute of Standards and Technology}
\acrodef{NDK}{Native Development Kit}
\acrodef{NTP}{Normal Temperature and Pressure}
\acrodef{OS}{Operating System}
\acrodef{PoP}{Package-on-Package}
\acrodef{OOE}{out-of-order execution}
\acrodef{PCB}{printed circuit board}
\acrodef{PID}{proportional-integral-derivative}
\acrodef{RAM}{Random Access Memory}
\acrodef{RISC}{Reduced Instruction Set Computing}
\acrodef{ROHC}{Robust Header Compression}
\acrodef{RMS}{Root Mean Square}
\acrodef{rpm}{revolutions per minute}
\acrodef{RTL}{Register Transfer Language}
\acrodef{RC}{resistor/capacitor}
\acrodef{RMSE}{root-mean-square error}
\acrodef{SIMD}{Single Instruction Multiple Data}
\acrodef{SMD}{surface mount device}
\acrodef{SoC}{Systems-on-Chip}
\acrodef{SGLP}{Super-graph Level Parallelism}
\acrodef{SLP}{Super-word Level Parallelism}
\acrodef{SPM}{Scratch-Pad Memory}
\acrodef{SVM}{State Vector Machine}
\acrodef{SRAM}{Static Random-access Memory}
\acrodef{SDRAM}{synchronous dynamic random access memory}
\acrodef{STP}{standard temperature and pressure}
\acrodef{TCP}{Transport Control Protocol}
\acrodef{TCT}{Task Completion Time}
\acrodef{TLB}{Translation Look-aside Buffer}
\acrodef{TLP}{Thread Level Parallelism}
\acrodef{TP}{Travaux Pratiques}
\acrodef{TMU}{Thermal Management Unit}
\acrodef{TTA}{Transport-Triggered Architecture}
\acrodef{UMTS}{Universal Mobile Telecommunications System}
\acrodef{VC}{Virtual Channel}
\acrodef{VM}{Virtual Machine}
\acrodef{VLSI}{Very-Large-Scale Integration}
\acrodef{VHDL}{VHSIC Hardware Description Language}
\acrodef{VLIW}{Very Long Instruction Word}
\acrodef{VM}{Virtual Machine}
\acrodef{WFL}{Weber-Fechner law}
\acrodef{WiFi}{Wireless-Fidelity}
\acrodef{WLAN}{Wireless Local Area Network}
\acrodef{WSN}{Wireless Sensor Network}

\newcommand{\rr}{\alpha}

\newcommand{\oa}{\omega_1}
\newcommand{\ob}{\omega_2}

\newcommand{\dgr}{{$^\circ$C}}

\newcommand{\dd}{\mathrm{d}}

\newcommand{\hac}{h_\mathrm{ac}}
\newcommand{\hpc}{h_\mathrm{pc}}

\newcommand{\diff}{d}

\usepackage{flushend}

\selectcolormodel{gray}

\begin{document}
%
\title{\color{black}Theoretical Analysis of Radiative Cooling for Mobile and Embedded Systems}
%
%
%
%

\author{Karel~De\,Vogeleer,
        Pierre~Jouvelot,
        and~Gerard~Memmi
\IEEEcompsocitemizethanks{\IEEEcompsocthanksitem K. De\,Vogeleer and Gerard Memmi are with TELECOM ParisTech, Universit\'e Paris-Saclay -- Deptartment INFRES -- CNRS LTCI - UMR 5141 -- Paris, France,\protect\\
Email: \{karel.devogeleer,gerard.memmi\}@telecom-paristech.fr
\IEEEcompsocthanksitem Pierre Jouvelot is with MINES ParisTech, PSL Research University, France. Email: pierre.jouvelot@mines-paristech.fr}
}

\IEEEtitleabstractindextext{%
\begin{abstract}
A new global analytical model of the heat dissipation process that occurs in passively-cooled embedded systems is introduced, and we explicit under what circumstances the traditional assumption that exponential cooling laws apply in such context is valid.
Since the power consumption and reliability of microprocessors are highly dependent on temperature, management units need accurate thermal models.
Exponential cooling models are justified for actively-cooled systems. Here, we analyze the tractability of the cooling law for a passively cooled body, subject to radiative and convective cooling, including internal heat generation.
Focusing then on embedded system-like objects, we compare the performance difference between our new passive cooling law and the conventionally-used exponential one.
We show that, for quasi isothermal cooling surfaces of the order of 1\,dm$^2$ or greater, the radiative cooling effect may become comparable to the convective cooling one.
In other words, radiation becomes non-negligible for systems with a cooling surface larger than about 1\,dm$^2$.
Otherwise for surfaces below 1\,dm$^2$, we show that the differences between the exact solution and the exponential cooling law becomes negligible.
In the absence of accurate temperature measurements, an exponential cooling model is shown to be accurate enough for systems, such as small-sized SoCs, that require low processing overhead.
\end{abstract}

}

\maketitle

\IEEEdisplaynontitleabstractindextext

%
\IEEEpeerreviewmaketitle

\ifCLASSOPTIONcompsoc
\IEEEraisesectionheading{\section{Introduction}\label{sec:introduction}}
\else
\section{Introduction}
\label{sec:introduction}
\fi

%
%
%
%


\IEEEPARstart{G}{iven} the crucial aspect of energy optimization in embedded and mobile systems, even a tiny amount of energy gained via a better understanding of thermal effects may have significant business and ecological impacts.
Temperature is an important factor influencing energy consumption of {\color{black} entire systems and, in particular,} microprocessors while executing programs.
Understanding and accurately modeling this relationship may bear impact beyond optimized system operation management.
This point is particularly acute for any system running on electrical battery such as mobile devices or sensors which participate in the \ac{IoT}.

Moreover, temperature and its variations affect the reliability of {\color{black} electronic} circuits.
Thermal gradients that occur both in space and time, induced by the variability in {\color{black}heat sources, e.g.,} microprocessor load and operations, generate thermal cycles that have an adverse affect on the failure rate of the system~\cite{Kong:2012:RTM:2187671.2187675}.
For example, a 10$^\circ$C to 15$^\circ$C temperature increase may halve a microprocessor's lifetime~\cite{Viswanath:2000:TPC}.
The \ac{ITRS} even states that processor costs and performance specifications may be limited by the lifetime reliability and is of primary concern in the microprocessor's design phase~\cite{5763032}.
Since power consumption increases exponentially with increasing silicon temperature~\cite{2014:devogeleer:samos}, thermal management techniques are employed to avoid self-destruction, to increase the \ac{MTTF} and minimize power consumption.
{\color{black}Moreover, from a user experience point of view, the skin temperature of portable devices should also be limited.
Experimental data show that the maximum skin temperature of hand-held devices should not exceed 41\dgr\ to 45\dgr, depending on the material, to assure the user's touch comfort~\cite{Berhe200733873}.}

Thermal management techniques may be deployed at the system design phase or can be deployed dynamically at run time by \acp{TMU} and \ac{DTM} systems.
A plethora of thermal control methods for microprocessors {\color{black}and embedded systems exist.
These methods} show trade-offs between temperature profile, frequency settings, power consumption and implementation complexity~\cite{Zanini:2013:OTC:2390191.2390197}.

Thermal management methods often incorporate a model describing the temporal thermal behavior of the {\color{black}system}.
Exponential-based models are popular, and scientifically sound for systems without internal heat generation and subject to active cooling, e.g., forced air or water cooling.
{\color{black}Exponential thermal behavior is also assumed in finite element analysis, as thermal capacities show \small{RC}-like behavior~\cite{1650228}.}
However, passively cooled systems, as frequently found in embedded systems, particularly mobile devices but also flats screen TVs etc., are not always forcibly cooled.
These {\color{black}passive systems} are subject to the same physical laws for dissipating their heat to the environment, but rely on different aspects of the heat dissipation process, {\color{black}such as radiative cooling.
Henceforth, passive cooling will imply the presence of radiative cooling.}

In this paper, we develop an accurate analytical solution to the problem of passive cooling of embedded systems.
It is important to understand the difference between an exponential cooling law and the cooling law of passively cooled devices since, in the literature, the radiative cooling aspect is frequently neglected.
We believe that this is because it is considered a secondary order factor and because of its non-linear nature, which poses problems in mathematical derivations and simulations.
In the case of active cooling, convective heat transfer {\color{black}usually} dominates the other heat transfer modes whereas, for passive cooling, radiation may become equally important, sometimes even more important, and may dominate the convective heat transfer mode.
{\color{black}This is especially true for large cooling surface areas, in the context of embedded systems.
Wagner and Maltz~\cite{Wagner5454212} also noted that the importance of radiation in dissipating the heat from exposed surfaces should not be underestimated.}
When radiation cannot be neglected, the transient thermal behavior of the system will deviate from an exponential cooling law.
In this paper we analyze under which circumstances the radiation is significant enough for it not to be neglected.
We show that the size of the cooling surface plays an important role in this question.
In particular, in cases where the cooling surface of the device is larger than about $> 1$\,dm${}^2$, the difference between the usual exponential model and passive cooling is significant.
Based on the passive cooling law's complex formulation, and in the absence of accurate temperature measurement samples, our work therefore suggests that an exponential cooling law is accurate enough for small {\color{black}systems, e.g., {\small SoC} applications,} and for systems that require low processing overhead.

{\color{black}We compare active and passive cooling processes of a system in the context of a mobile embedded device, i.e., a computer system including internal heat generators and subject to cooling.}
The main contributions of this paper are:
\begin{itemize}
 \item the accurate analytical solution for the problem of (passive) cooling of a system subject to radiation, convection, and internal heat generation;
 \item approximations to the exact analytical solution for use in practical \acp{DTM} of embedded systems, validated by intense simulations;
 \item actionable rules-of-thumb to decide when passive cooling becomes non-negligible compared to active cooling in embedded systems.
\end{itemize}

The rest of the document is developed as follows.
Section \ref{sec:Thermal_Management_Techniques} highlights the use of cooling laws in existing research related to thermal management units {\color{black} in embedded applications}.
Section~\ref{sec:coolinglaws} develops the exact cooling law for microprocessors subject to passive cooling;
this law is also validated via finite-element simulations and approximations are analyzed.
Besides, the impact of active cooling of microprocessors is also discussed.
{\color{black}Section~\ref{sec:laws_differ} studies the performance difference between the exponential cooling law and the passive cooling law, based on our analytical model.}
We conclude in Section~\ref{sec:conclusion} with a summary and give directions for future research.


\section{Radiative Cooling in Existing Thermal Management Techniques}
\label{sec:Thermal_Management_Techniques}

Thermal management techniques for embedded systems have been devised to control their heat dissipation.
Excessive heat dissipation may have adverse effects on performance, the short term and long term failure rate of the {\color{black}system, and microprocessors in particular.}
Basic run-time thermal management decisions can be rudimentary, such as {\color{black}using smart sleep modes} or clock gating.
Yet, if service continuation is needed, more advanced thermal techniques are required.
Thermal-aware design of systems and microprocessors can also be effective to minimize peak and average heat dissipation during run time.
The challenge here, however, lies in decision making based on incomplete design and run-time detail information.

To get a current perspective on how such issues are addressed in the literature, we surveyed top computer architecture and \ac{VLSI} conferences for papers devoted to \acp{TMU}, \acp{DTM} and temperature-aware design methods based on heat transfer theory.
The conferences surveyed are {\small ISCA}, {\small MICRO}, {\small ASPLOS}, {\small HPCA}, {\small PACT}, {\small ISLPED}, {\small ICCAD}, {\small DAC}, {\small DATE}, {\small ASP-DAC} from 2010 to 2014.
We identified 35 papers focusing on the thermal optimization of microprocessors or embedded systems using heat transfer models.
90\% of these papers base their results solely upon simulation or numerical analysis; the remaining ones use either actual measurements or a combination of simulation and measurements to make their point.
Beside custom thermal simulators and models, non-commercial and open-source thermal simulators are mostly used: these are based on finite-element methodologies.
Commercial applications such as {\small COMSOL} Multiphysics$^\circledR$, Autodesk Simulation CFD or {\small FLoTHERM}$^\circledR$, which support the radiative heat transfer mode, are not used in the selected papers.
About 40\% of the selected papers deploy Hotspot for their thermal simulations.
Hotspot~\cite{1650228} is a self-proclaimed accurate and fast thermal model designed for microprocessor architectural analysis, e.g., floor planning.
The basic setup of Hotspot includes active cooling via a heat sink.
No passive cooling capabilities are available in Hotspot.
Other experimental simulators, such as LightSim~\cite{6742997}, {\small CONTILTS}~\cite{DBLP:journals/jolpe/HanKK07}, {\small ISAC}~\cite{4039519} and PowerBlurr~\cite{5444285}, also allow for thermal analysis of microprocessors, but are less popular and again, none support radiative cooling.
In most of the simulations, the temperature at steady-state and transient temperatures are available, where the steady-state case is much faster to compute than the transient behavior.

It is worthwhile to ponder upon why no non-commercial simulators support radiative cooling.
One reason could be that the non-linear behavior of radiation is not easy to handle in mathematical formulations although advanced finite element techniques could be employed in numerical simulations.
Also, it is not always clear to what extent radiation actually affects the thermal behavior {\color{black}of semiconductors or embedded systems.}
As a result, given the lack of passive cooling capabilities in many simulators, it is not surprising that passive cooling has not gotten much attention in the thermal management research community.
In fact, we found only one paper~\cite{Vincenzi:2011:FTS:2016802.2016842}, about 3D integrated circuits, which mentions that radiation may influences the thermal behavior of microprocessors; yet in this work no further reference to radiation is found.
Nevertheless, 30\% of the papers we surveyed claim that their research is applicable to mobile embedded systems, a situation in which passive cooling is usually of the essence.

Beside generic thermal microprocessor simulators, dedicated embedded system thermal simulators were also developed.
Therminator~\cite{Xie:2014:TTS:2627369.2627641}, for example, is a thermal simulator designed to simulate heat dissipation in smartphones.
Finite element methodologies are used to compute the heat propagation through an arbitrary {\color{black}heterogeneous} smartphone configuration, which includes a \ac{PCB}, battery, case, display etc.
The authors show that their dedicated thermal simulator produces results that are close to what commercial \ac{CFD} software would calculate.
Therminator takes the convective and conduction heat transfer modes into account.
Again, heat loss via radiation, however, is not implemented in their thermal simulator.
Luon et al.~\cite{Luo20081889} analyzed the issue of thermal management on mobile phones based on numerical simulation and basic thermal models.
The authors came up with design proposals on how to improve the thermal management of mobile phones by studying the steady-state behavior of the system.
Even though radiation is mentioned in the introduction including formulations, radiation is not present in their stead-state analysis.
Gurrum et al.~\cite{6249032} decomposed, just as Luo et al.~\cite{Luo200
81889}, a hand-held device in multiple subparts with different physical properties and analyzed its thermal behavior.
Radiation, however, did not come to their attention.
{\color{black}Lee et al.~\cite{4544324} modeled the steady-state thermal behavior of hand-held  electronic devices using \small{ANSYS}, a commercial finite element simulator.
The authors enabled radiation in their simulations.
However, they do not discuss to what extend their results are affected by the presence of radiation.
The data they published do not allow to estimate its impact either.}

From our literature survey we conclude that the numerical tools used for thermal behavior of embedded systems can be classified into three categories.
First, we have the general-purpose \ac{CFD} software, which is able to simulate arbitrary systems including all modes of heat transfer.
These systems require the most efforts to produce interesting results.
The second class corresponds to dedicated embedded system simulators.
We have observed that the designers of the simulators are aware of surface radiation but they do not provide support in their simulators.
And last, which are the most popular, are the generic microprocessor thermal simulators.
We have not seen any of these microprocessor simulators supporting the radiative heat transfer mode.

{\color{black}This state of affairs provides us with a strong motivation for our work to go beyond previously-established thermal models by incorporating radiative cooling capabilities.
Our work strives to understand the possible impact of radiation on the transient and steady-state thermal behaviors of microprocessors in the context of embedded systems.}

\section{Cooling Laws}
\label{sec:coolinglaws}

The exponential cooling law is the most widely used cooling law to model the thermal behavior of {\color{black} entire embedded systems or microprocessors}, as shown by our literature survey.
The rationale behind an exponential law is based on temperature traces of forcibly cooled systems, which indeed show clear exponential behavior~\cite{Luo20081889,6249032}.
One may attribute the exponential curve to Newton's law of cooling.
However, the presence of internal heat generation, which renders the direct applicability of Newton's law of cooling irrelevant for computer systems, should not be forgotten.
In the sequel we show however that Newton's law of cooling extended with internal heat generation also yields an exponential cooling law.
For passively cooled microprocessors, the radiative heat transfer mode, beside natural convection, also needs to be taken into account.

In this section, after a brief overview of basic heat transfer principles~\cite{cengel2010heat}, we develop the cooling law for an actively cooled {\color{black}system with internal heat generation}.
We then adept this model to radiative cooling to obtain our first contribution, a representative model for passively cooled systems {\color{black}with internal heat generation.
Besson~\cite{0143-0807-31-5-013} used the same approach to model radiative cooling.
Besson showed, by comparing experimental data with his analytical results, that this approach is adequate in modeling thermal-related physical problems.}

\subsection{Basics of Heat Transfer}

Heat transfer happens via a combination of the three fundamental modes: \emph{convection}, \emph{conduction}, and \emph{radiation}.
Each of these modes follows its respective law.
In the sequel we assume an isothermal body {\color{black}with internal heat generation} that cools via convection and radiation.
Isothermal conditions may be approximated if the body heats up uniformly, or if the internal heat conduction happens considerably faster than the heat loss of the body to the environment.
Therefore we won't discuss conduction in detail.

A solid body immersed in a moving fluid, e.g, air or water, is subject to energy exchange if the temperatures of the body and the moving fluid differ.
Energy is \emph{convected} from or to the body if the moving fluid has a different temperature from the body.
The energy transfer rate $q$ [W] between the moving fluid and the surface of the body is formally known as \emph{Newton's law of cooling}:
\begin{equation}
 q = C \frac{\diff T}{\diff t}= \hac S (T_m - T),\label{eq:netwoncooling}
\end{equation}
where $T_m$ is the temperature of the moving fluid (environment), $S$, the cooling surface area of the body, and $\hac$, the \emph{convective heat transfer coefficient} [W/(m$^2\cdot$K)].

Radiative heat transfer happens through exchange of electromagnetic waves, possible through both vacuum and transparent media.
Stefan-Boltzmann's law states that the power radiated from a \emph{blackbody} is proportional to its temperature.
A \emph{blackbody} is a body that absorbs all incident radiation.
In particular, Stefan-Boltzmann's law states that the radiative heat transfer rate $q$ is proportional to the blackbody's temperature to the 4th power:
\begin{equation}
q = \epsilon\sigma S T^4,\label{eq:Stefan-Boltzmann-law}
\end{equation}
where $\epsilon\in[0,1]$ is the \emph{emissivity} of a gray body's surface (dimensionless), and $\sigma$ is the Boltzmann constant $5.6697 \times 10^{-8}$ [W/(m$^2\cdot$K$^4$)].
A \emph{gray body} is a body that reflects a certain amount of the incident radiation.
The emission and absorption of a gray body can be well represented by a blackbody's behavior scaled by its emissivity: $0 \leq \epsilon \leq 1$.
In practical situations the total heat loss of a body via radiation is equal the emitted radiation minus the absorbed radiation:
\begin{equation}
q = \epsilon\sigma S (T^4_a - T^4),
\end{equation}
where $T_a$ is the radiation temperature of the environment.
Here we implicitly assumed that the environment has the same emissivity as the body itself.


The total heat transfer from a body happens via the combination of the basic heat transfer modes.
Beside, a body may also produce heat $H(\cdot)$ [W] which is referred to as \emph{internal heat generation}.
The internal heat generation may be a function of space, time, temperature or others.
{\color{black}In the sequel we will assume that the internal heat generation is homogeneously present throughout the entire body, independent of time, but dependent on temperature}.


\subsection{On the Isothermal Assumption}

\begin{figure*}[!t]
  \centering
  \hfill
  \subfloat[an Apple iPad by Wagner and Maltz~\cite{Wagner33873}]{  \includegraphics[width=0.49\linewidth]{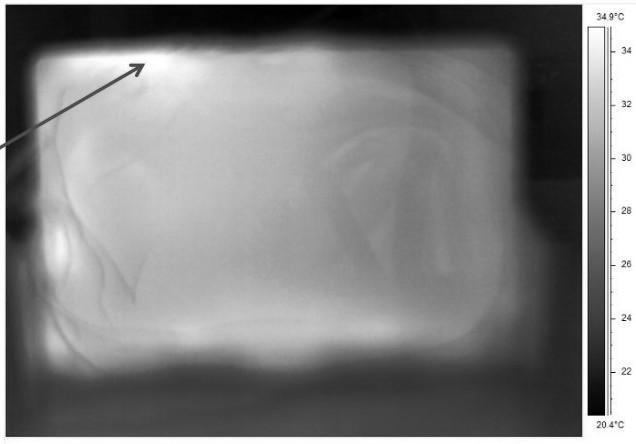}
  \label{fig:appleiPad}}
  \hfill
  \subfloat[an unnamed thin notebook by Mongia et al.~\cite{Mongia2008992}]{  \includegraphics[width=0.45\linewidth]{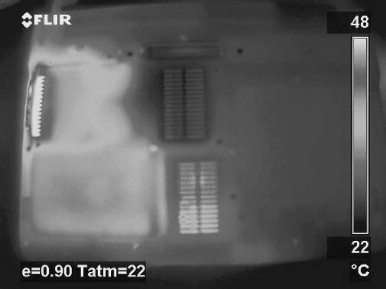}
  \label{fig:laptop}}
  \hfill
  \caption{\color{black}Experimental thermal imaging of the skin temperature of (a) an Apple iPad~\cite{Wagner33873} and (b) an unnamed thin notebook~\cite{Mongia2008992}. The iPad exemplifies the quasi-isothermal surface of an embedded system. The surface temperature varies between 30\dgr\ and 35\dgr. On the other hand, the thin notebook shows large temperature variations, between 25\dgr\ and 48\dgr. Here, aggressive active cooling methods extract the heat as fast as possible from the heat sources inside the device.}
  \label{fig:thermal_imaging}
\end{figure*}

{\color{black}Our work assumes quasi-isothermal conditions of the system under study, meaning that the temperature is quasi-constant throughout the surface of the system.
Assuming quasi-isothermal conditions simplifies the mathematical derivation of the transient thermal behavior considerably, as we will see further.
We can observe isothermal conditions of embedded systems in practice.
For example, in Figure~\ref{fig:appleiPad}, Wagner and Maltz~\cite{Wagner33873} showed via thermal imaging that the surface of an Apple iPad has near-isothermal properties.}
{\color{black}The reason why this tablet shows a quasi-isothermal profile is that inside the tablet a fan driving a copper heat duct is installed to distribute internally the heat generated by the \ac{SoC}.
This results in a more or less uniform thermal profile for the tablet.
Besides active heat distribution techniques, polymer/graphite/copper/aluminum-based heat spreaders are also often used to facilitate passive head spreading within a computer system.

Isothermal properties may, however, not always be present in embedded systems.
Figure~\ref{fig:laptop} shows an example of the thermal image of an unnamed thin and light notebook~\cite{Mongia2008992}.
In this example, it is evident that the thermal profile can less likely be deemed isothermal within the thermal operating range of an embedded system.
Acquiring a detailed thermal profile of a three-dimensional system is perhaps almost impossible to obtain.
The trade-off between tractability and accuracy has to be addressed when it comes to thermal profiling of an embedded system.
Temperature sensors can be installed to measure the temperature at certain spots, but then the space between sensors is not covered and must be interpolated.
Thermal imaging, as shown in Figure~\ref{fig:thermal_imaging}, shows a detailed thermal profile, but only of the surface of the system.
From an analytical point of view, equations were developed that express the transient and steady-state thermal behaviors of systems subject to convective cooling.
One of the most realistic analytical modeling was done by Yovanovich~\cite{693656884} and Lee et al.~\cite{ssdf5sd5466s8d74} for non-isothermal axisymmetric cylindrical homogeneous bodies subject to convection and a heat source applied to one side.
These analytical expressions are, however, rather complex and tend to be expressed in a non-closed form.
Furthermore, convection has linear properties w.r.t. temperature; radiation, on the other hand, shows non-linear properties.
Thus, adding support for radiative cooling may render those analytical derivations even more complex.
An alternative to analytical expressions to obtain non-isothermal profiles of systems is to resort to \ac{CFD} or finite element simulations.
However, neither \ac{CFD} simulations nor non-closed form equations are tractable for online thermal optimization methods with limited resources, in terms of performance and energy, such as those found in embedded system.
In such situations, assuming quasi-isothermal conditions may be an effective method to trade overhead for a sufficient level of accuracy.
Moreover, in a system sporting only one temperature sensor, one doesn't have many alternatives but applying Occam's razor principle and assuming isothermal properties while assessing the thermal behavior.


In the sequel, we will assume isothermal behavior to estimate the magnitude of the radiative cooling component, aiming at assessing its importance compared to other cooling modes.
This approach is based on our closed-form analytical equation, which can easily be applied to other applications.
This is not meant to be a detailed modeling, but rather a method to measure the influence of the radiative cooling component and a first-order approximation of the transient thermal behavior of a computer system.
For dealing with the case where a non-isothermal profile is key, we advise to look into complex tools such as finite element simulations.
}

\subsection{Active Cooling: the Newtonian Approach}
\label{sec:newtonapproach}

{\color{black}Actively cooled systems spend energy to forcibly cool down the system.
The most basic and widely used active cooling technique is an air fan mounted directly on the system, or on a heat sink attached to the system.
More advanced actively cooled systems include fluid cooling.
Fluids-based cooling devices are more effective but also more expensive, more complex to maintain and more hazardous for the hardware.
Examples of technologies under development for active thermal management of portable electronic devices are phase-change materials, micro heat pipes, conductivity materials such as carbon~\cite{Grimes20102363}, thermoelectric cooling, and two-phase refrigerant cooling.

Active cooling is usually associated with Newton's law of cooling.}
Newton's law of cooling states that the temperature rate of change of a system is proportional to the difference between the ambient temperature and the system's temperature.
\acfp{TMU} and \acfp{DTM} often assume the system to cool down following Newton's law of cooling.
{\color{black}Newton developed his law experimentally for systems under the following conditions:%
\begin{enumerate}
	\item the body is quasi-isothermal throughout;
	\item it conducts heat much faster than it gains from the surrounding; and
	\item the body's average temperature is not too large.
\end{enumerate}%
The latter condition implies the neglect of radiation.
Gockenbach and Schmidtke~\cite{Gockenbach:netwonheat} showed analytically, via heat transfer theory, that under these conditions indeed the cooling process can be approximated by an exponential-based law satisfactorily.
Newton's conditions are frequently assumed in experimental thermal management systems~\cite{Kong:2012:RTM:2187671.2187675,10.1109/L-CA.2003.5,Zhang:2007:AAT:1326073.1326131,10.1109/TPDS.2007.1092,conf/asplos/HeathCGRJ06,Jayaseelan:2008:TAT:1509456.1509593,Forte:2013:ETV:2465787.2465798}.
For actively cooled systems an exponential assumption is a good approximation when radiative and conductive cooling may be neglected, as we explain in the sequel.

Let's take a look at an actively cooled system with an internal heat source.}
Assume that for an isothermal system the stored energy is approximated by the sum of the heat transfer induced by \emph{convective cooling}: $\hac S (T_m-T)$, and an \emph{internal heat generation} (ihg): {\color{black}$\eta_1 T + \eta_0$, which we deem linearly temperature-dependent} as a first-order approximation:
\begin{eqnarray}
C \frac{\diff T}{\diff t} & = & \text{convection} + \text{internal heat generation} \nonumber\\
& = &  \text{$\hac S(T_m -  T)$} + (\eta_1 T + \eta_0).\label{eq:start1}
\end{eqnarray}
where $C$ is the body's heat capacity and $T_m$ the ambient temperature.
Note that, if the active cooling system consists of a fan and heat sink, then $\hac$ depends upon the dimensions of the heat sink, and the \ac{rpm} of the fan.
Moreover, $\eta_1$ and $\eta_0$ are also dependent on the activity and the temperature-dependency of the heat source.
{\color{black}For example, for microprocessors, the clock frequency, type of computations, and load on the system, or the brightness of an \small{LCD} display, could affect the heat generation.
Then,} similar to Weissel and Bellosa's~\cite{weissel04thermalmanagement} work, one gets, from Equation~\ref{eq:start1}:
\begin{eqnarray}
T - \frac{\eta_0+\hac S T_m}{\hac S-\eta_1} & = & c_0 e^{-\frac{(\hac S-\eta_1) }{C} t},
\end{eqnarray}
while imposing the initial condition at $t=0$: $T(0)=T_0$.
Therefore $c_0=T_0-\frac{\eta_0+\hac S T_m}{\hac S-\eta_1}$, and thence
\begin{equation}
T_\mathrm{ac}(t) = \frac{\eta_0+\hac S T_m}{\hac S-\eta_1} + \left(T_0-\frac{\eta_0+\hac S T_m}{\hac S-\eta_1}\right)e^{-\frac{(\hac S-\eta_1) }{C} t}.\label{eq:solution-newton}
\end{equation}
It is clear that such a system is only stable if the cooling process with constant $\hac$ convects heat away from the system faster than the system is generating internal heat.
The system is stable if there exists an equilibrium temperature $T_e$ for the system, which is equivalent to saying that
\begin{equation}
0  = \hac S(T_m - T_e) + (\eta_1 T_e + \eta_0)  \Rightarrow  \hac  = \frac{\eta_1 T_e + \eta_0}{S(T_e - T_m)},\label{eq:hatTe-active}
\end{equation}
where all constants $\{T_e,T_m,\eta_1,\eta_0\}\in\mathbb{R}^+$.
We can state, given that $\hac$ must be positive, that $T_e > T_m$.
We can also conclude from Equation~\ref{eq:solution-newton} that $\hac$ is always larger than $\eta_1/S$.
If $\hac < \eta_1/S$, the exponent in Equation \ref{eq:solution-newton} would go to infinity over time.
In practical applications, the value of $\hac$ must be dimensioned properly such that the system's $T_e$ stays below the maximum operation temperature.

Not surprisingly, Newtonian cooling with linear internal heat generation yields again an exponential relationship between temperature and time.
Consequently, the power $P$ consumed by the system, which is an affine transformation of temperature ($ihg = \eta_1 T + \eta_0 $), will also exhibit exponential behavior.
An exponential model for actively cooled systems with linear (or constant, $\eta_1 = 0$) internal heat generation is therefore a valid approximation.
The exponential assumption is however not quite the same as assuming simple Newtonian cooling, as the coefficients in both models are different, mainly due to the presence of the internal heat generation.
In the case of the presence of internal heat generation, the equilibrium temperature $T_e$ of the system will be larger than the ambient temperature, see Equation~\ref{eq:solution-newton} for $t \rightarrow \infty$.


\subsection{Passive Cooling via Radiation, (Natural) Convection and subject to Internal Heat Generation}
\label{sec:radcooling}

We now adapt the previous model, {\color{black}designed for for active cooling}, to better fit passively cooled embedded systems.  
Systems that are not actively cooled must indeed rely on passive cooling to attain a temperature equilibrium state.
Passive cooling mechanisms include radiation, but also natural convection.
Note though, that convection may be considerably smaller than when the system is actively cooled.
The convection arising here may be originating from buoyancy forces, or natural movement of air, e.g., wind.
In the case of buoyancy forces, sometimes the convection is referred to as \emph{natural convection} as the movement of air is not enforced on the system.

Let's assume an isothermal body subject to radiative cooling and convection with internal heat generation.
The temperature change of such an object at any given point in time is equal to the heat absorbed from the environment, plus the internal heat generation, minus the heat released to the environment.
Absorption of heat happens via radiation whereas the release of heat is happening both via radiation and convection.
The temperature change of such a system, with internal heat generation (ihg), can be represented by the following equation:
\begin{eqnarray}
	\frac{\diff T}{\diff t} & = & \frac{1}{C}(\text{radiation} + \text {convection} + \text{ihg}) \nonumber\\
               & = & {\epsilon \sigma S ( T^4_a - T^4)} + hS(T_a -  T) + (\eta_1 T + \eta_0),\label{eq:start}
\end{eqnarray}
where $\epsilon$ is the emissivity of the body, and $\sigma$ is the Boltzmann constant.
Here it is assumed that the internal heat generation is linearly dependent on the temperature of the body: $H(T)=\eta_1 T + \eta_0$. Yet, higher order polynomials (up to the 3rd order) can be used as well for the following derivation to hold (as shown in Appendix~A).
Also, $T_m$ is presumed to be equal to $T_a$.

By rearranging Equation~\ref{eq:start} we obtain:
\begin{equation}
\frac{\diff T}{\diff t} = \frac{1}{C} \{ - \epsilon \sigma S T^4  +  ( \eta_1 - hS) T + ( \eta_0 + S[ h T_a + \epsilon \sigma  T^4_a ])\}. \label{eq:diff-passive}
\end{equation}
Here, the right-hand side is a 4th-order polynomial.


The derivation (provided in Appendix~B) shows that the exact solution to the problem of cooling of a {\color{black}system} subject to radiation, convection, and internal heat generation is given by Equation~\ref{eq:solution-passive}.
\begin{multline}
 t = -\frac{1}{\kappa_4}\bigg(A\ln|T-\oa | + B\ln|T-\ob | + \frac{C}{2} \ln| (T-\rr)^2 \\
  + \beta^2| +  \frac{\alpha C-D}{\beta}\arctan\left( \frac{T-\rr}{\beta} \right) + c_o \bigg),\label{eq:solution-passive}
\end{multline}
Here, $c_o$ must satisfy the initial conditions $t(T_0)=0$, if $t(T)$ denotes the right-hand side expression in Formula~\ref{eq:solution-passive}:
\begin{multline}
c_o -A\ln|T_0-\oa | - \frac{C}{2} \ln| (T_0-\rr)^2 + \beta^2| \\
 - B\ln|T_0-\ob | - \frac{\alpha C+D}{\beta}\arctan\left( \frac{T_0-\rr}{\beta} \right),
\end{multline}
the $\omega_*$ are the roots of the 4th-order polynomial given in Equation~\ref{eq:diff-passive} (we define $\omega_{1,2}$ as the real roots, $\omega_{3,4}$ as the complex conjugates), and
\begin{subequations}
\begin{empheq}{align}
   A  & =~ \frac{1}{ (\oa-\ob)((\Re(\omega_3)^2+\Im(\omega_3)^2)-\oa (2\Re(\omega_3)-\oa))} \nonumber\\
   B  & =~ -A \frac{\Re(\omega_3)^2 + \Im(\omega_3)^2 - \oa (2\Re(\omega_3)-\oa) }{\Re(\omega_3)^2 + \Im(\omega_3)^2  -  \ob (2\Re(\omega_3)- \ob) }   \nonumber\\
   C  & =~ -(A+B) \nonumber\\
   D  & =~ A (2\Re(\omega_3)-\oa) + B (2\Re(\omega_3)- \ob),\nonumber
\end{empheq}
\end{subequations}
where  $\Re$ and $\Im$ denote the real and imaginary parts of complex numbers, respectively.

Surprisingly, our result is consistent with the solution presented by Besson~\cite{0143-0807-31-5-013}, even though he modeled a different physical problem.
Besson however assumed some simplifications, different assumptions from ours, and solved the differential equation via other methods.
Nonetheless his solution also contains three \emph{logarithms}, one of them containing a second-order polynomial, and an \emph{arctan}.
Because of Besson's simplifying assumptions, however, his equation is limited to the case where $T-T_a = T+\frac{\eta_1}{\eta_0}$, which is a special case of our initial problem.

Similarly to actively cooled system, the passively cooled system will tend towards an equilibrium temperature $T_e$ only if Equation~\ref{eq:diff-passive} equates to zero.
Given that an equilibrium temperature $T_e$ exists, the convective heat transfer coefficient $\hpc$ must be such that
\begin{eqnarray}
\hpc  & = & \frac{\eta_1 T_e +  \eta_0 + \epsilon\sigma S (T^4_a-T_e^4) }{S(T_e - T_a)},\label{eq:hatTe-passive}
\end{eqnarray}
where all constants $\{T_e,T_a,e,S,\eta_1,\eta_0\}\in\mathbb{R}^+$.
Consequently, this is only possible if $ T_e > T_a$, as in the case of active cooling, and $\eta_1 T_e + \eta_0 > \epsilon \sigma S (T_a^4 - T_e^4)$.

The accurate solution for passively cooled objects as presented in Equation~\ref{eq:solution-passive} {\color{black}yields time in} function of the temperature: $t(T)$.
For practical reasons, such as for \acp{DTM}, \acp{TMU}, or \ac{PID} control techniques, an analytical formulation in the form of $T(t)$ is preferred.
Inverting the exact solution is however, not a straightforward task, mainly because the \emph{arctan} is hard to deal with as it keeps recurring.
Numerical approaches will thus be preferred to compute this exact inverse solution.
In Section~\ref{sec:approx_f(T)_t} we will discuss approximations to the exact solution.
{\color{black}For the interested reader, more details of the passive cooling law given by Equation~\ref{eq:solution-passive} is developed by De\,Vogeleer~\cite{kdvPHDthesis}.}

\subsection{Experimental Validation of the Accurate Cooling Law Applied to a Microprocessor}
\label{sec:experimental_validation}

To validate the passive cooling solution defined in Equation~\ref{eq:solution-passive}, we setup a set of \ac{CFD} simulations in {\small COMSOL} where we analyze the transient thermal behavior of a slice of silica glass (SiO$_2$), as silica glass is close to the thermal properties of a microprocessor.
A {\small 3D} conjugate heat transfer scenario was created, with simulation settings as shown in Table~\ref{table:comsol_settings}.
\begin{table}
   \centering
   \caption{{\color{black}Configuration of the {\smaller COMSOL} simulations used for the validation of our analytical model}. Specific values were calculated for the convective heat transfer coefficient ($\hac$) and internal heat generation (ihg) such that a predefined equilibrium temperature is reached.}\label{table:comsol_settings}
   \begin{tabular}{|c|c|c|} \hline
     \multicolumn{3}{|c|}{\sc Constants} \\\hline
     symbol & value & dim. \\\hline
     $\sigma$   & 5.670$\times$ 10$^{-8}$     & W/(m$^{2}\cdot$K$^{4}$)  \\
     $\epsilon$ & 0.94 & -     \\
     $T_a$ & 20 & $^\circ$C    \\
     $D$   & 2  & mm           \\
     $S$   & 0.01 & m$^2$       \\
     $C$   & $S\times D\,\times$\,1548709 &   J/K\\\hline
     \multicolumn{3}{|c|}{\sc Variables} \\\hline
     symbol & value & dim. \\\hline
     heating: $\hac$ & 11.144 & W/(m$^2\cdot$K) \\
     cooling: $\hac$ & 76.939 &  W/(m$^2\cdot$K) \\
     $\eta_1$ & 9.407  & W/K $\times$ 10$^{-3}$ \\
     $\eta_2$ &  1.318 & W \\\hline
   \end{tabular}
\end{table}  
The exact same values, as listed in this table, were also used in our theoretical model.
To approximate an isothermal object in {\small COMSOL} we have multiplied the thermal conductivity of the silica glass by $10^3$, {\color{black}in a way similar to Wagner and Maltz's approach~\cite{Wagner5454212}}.
The silica glass has a surface area of 0.01\,m$^2$.
For the heating process $T_0$ is set to 25$^\circ$C and $T_e$ is scaled between $T_0$ and 45$^\circ$C.
Similarly, for the cooling process $T_0$=45$^\circ$C and $T_e$ is scaled between $T_0$ and 25$^\circ$C.
The temperature values we chose correspond to what is typically encountered when using a mobile device.
We used linear internal heat generation with the parameters as shown in Table~\ref{table:comsol_settings}.
The convective heat transfer coefficient $\hac$ was set (Equation~\ref{eq:hatTe-passive}) such that with the given internal heat generation the predefined equilibrium temperature is attained.
We look at levels of internal heat conversion derived from ARM Cortex A15 quad-core processor power measurements on the Exynos 5210 \ac{SoC}~\cite{2014:devogeleer:samos}.
The shown internal heat conversion represents the A15 processor running at maximum frequency while executing four applications in parallel.

\begin{figure}[!t]
  \centering
  \input{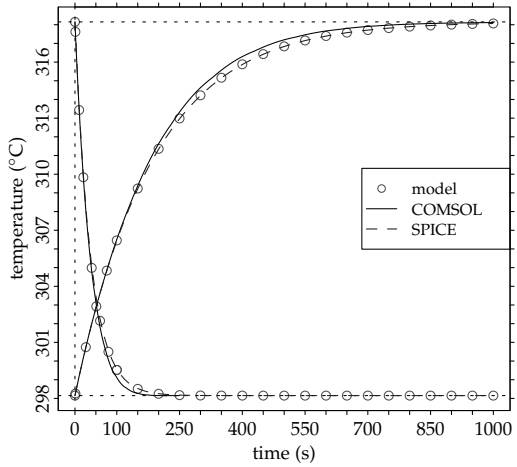}
  \caption{\color{black}A realistic example of the transient thermal behavior as per {COMSOL}, SPICE and the analytical cooling model from Section~\ref{sec:radcooling}. The parameters, as shown in Table~\ref{table:comsol_settings}, were used to simulate the cooling of a tablet-like object with internal heat generation representative for a powerful SoC microprocessor. The transient thermal behavior and errors for different levels of internal heat generation and surface size look similar.}
  \label{fig:comsol-tran}
\end{figure}

Figure~\ref{fig:comsol-tran} shows the transient thermal behavior of the silica glass as described above.
Both the cooling and heating process are shown in the same graph.
We have also generated data for various surface areas, internal heat generation levels and equilibrium temperatures; since all graphs look similar we don't show all of them.
Our theoretical model curves follow the experimental {\small COMSOL} curves well.
The maximum temperature difference between our model and the {\small COMSOL} results is less than  0.5$^\circ$C.
Interestingly, the {\small COMSOL} transient data seems to have a slightly steeper slope than our theoretical model.
This could be originating from the fact that the {\small COMSOL} object is not 100\% isothermal.
{\color{black}Figure~\ref{fig:comsol-tran} also shows the results of a simulated electrical circuit in \small{SPICE}, modeling the same cooling problem, based on the current/thermal equivalence~\cite{azar1997thermal}.
The temperature dependency of the radiative component was modeled with a \emph{voltage-controlled current source} to simulate its non-linear properties.
The \small{SPICE} simulations follow the analytical results systematically well.
The maximum difference is around 25\,mK, which is negligibly small.}

Despite the small temperature discrepancy between our analytical model, the {\small COMSOL} data {\small SPICE} we may deem our model an appropriate solution for passive cooling with internal heat generation.

\subsection{Approximations of the Accurate Cooling Law}
\label{sec:approx_f(T)_t}

The accurate solution for the passive heat Equation~\ref{eq:solution-passive} is of the form $f(T)=t$.
Ideally, for practical motivations, we would like to know the inverse $f(t)=T$.
For example, this may be convenient for the equation to be used in \ac{PID} controller systems.
Calculating the inverse of Equation~\ref{eq:solution-passive} is, however, a challenging endeavor.
Therefore, we will utilize effective approximations to obtain an invertible heat equation.

Finding a useful expression $f(t)=T$ requires isolating $T$ in Equation~\ref{eq:solution-passive}.
Mainly the presence of the \emph{arctan} {\color{black}challenges} the mathematical derivation.
Linearization or differential approximation will not provide any help as the derivative within the pertinent temperature range, i.e., between 25$^\circ$C and 45$^\circ$C, is far from being constant.
Converting the \emph{arctan} into a logarithm introduces imaginary numbers; yet, applying complex exponentiation rules will not get rid of the \emph{arctan}.
The \emph{arctan} keeps recurring further on in the derivation.
So we need to walk different paths to come to a solution for $f(t)=T$.

Table~\ref{table:approximations:overview} shows an overview of three different approximations that we will consider.
\begin{table}
   \renewcommand{\arraystretch}{2}
   \caption{\color{black}Summary of the presented approximations to the accurate passive cooling law. The \emph{coefficient approximation} approximates Stefan-Boltzmann's law with a quadratic polynomial. The \emph{O'Sullivan approximations} use binomial expansion to reduce the polynomial order of the cooling law. \label{table:approximations:overview}}
   \centering
   \begin{tabular}{|c|c|} \hline
     {\sc Approximation} & {\normalsize $T(t)$} \\\hline
     Coefficient & {\normalsize $T = \frac{\omega_1 \pm \omega_2c_o e^{-\frac{\kappa_2}{A} t}}{1\pm c_oe^{-\frac{\kappa_2}{A}t }}$} \\\hline
     O'Sullivan 1st & {\normalsize $T = \left(T_0-T_a + \frac{p}{n}\right) e^{-\frac{n}{C}t} - \frac{p}{n} + T_a$} \\\hline
     O'Sullivan 2nd & {\normalsize $T = \frac{\omega_1 \pm \omega_2c_o e^{-\frac{m}{A} t}}{1\pm c_oe^{-\frac{m}{A}t }}$} \\\hline
   \end{tabular}
\end{table} 
The derivation and motivation behind each approximation, as well as the definition of all the variables, are expounded in Appendix~C.
In short, the \emph{coefficient approximation} models the radiation within a specific temperature range with a quadratic polynomial.
This reduces Equation~\ref{eq:diff-passive} to a second-order problem.
The \emph{first} and \emph{second O'Sullivan approximations} are based on a binomial expansion~\cite{Sullivan:newton} that mingles the coefficients of Equation~\ref{eq:diff-passive} in a deterministic manner.
The advantage is that the resulting equation is invertible when higher-order coefficients are dropped.
Also, the accuracy of the approximation can be controlled by the degree of coefficients selected.
As can be observed from Table~\ref{table:approximations:overview} the coefficient approximation and the second-order O'Sullivan approximation are similar in shape.
However, the definition of their respective variables have no common ground.

Let us analyze the accuracy of the approximations.
We define the measure of accuracy as the \ac{RMSE} between the accurate cooling solution $\phi$ and an approximate solution $\psi$ for $n$ samples:
\begin{equation}
 \text{RMSE} = \sqrt{ \frac{\sum_{i=0}^n(\phi_i-\psi_i)^2 }{n}},
\end{equation} 
where $n$ is the number of samples over which \ac{RMSE} is computed.
We define $n$=500 and equally spaced between $t\in\{0,t(0.99\cdot T_e)\}$ (see Equation~\ref{eq:solution-passive} for $f(T)=t$).
The accurate cooling law and its approximations are generated with the same constants as the {\sc\small COMSOL} simulation of the previous section in Table~\ref{table:comsol_settings}.
We investigate the accuracy while changing surface area $S$, internal heat generation (ihg),
 equilibrium temperature $T_e$, for the cooling and warming process separately.
We set $T_0$=25$^\circ$C for the heating process and $T_0$=55$^\circ$C for the cooling process.
We variate the equilibrium temperature $T_e$ between 25$^\circ$C and 55$^\circ$C.
The convective heat transfer coefficient is computed accordingly to attain the respective equilibrium temperature based on Equation~\ref{eq:hatTe-passive}.
The variables generated for the accurate cooling law are then used to compute the approximations.

\begin{figure*}[!t]
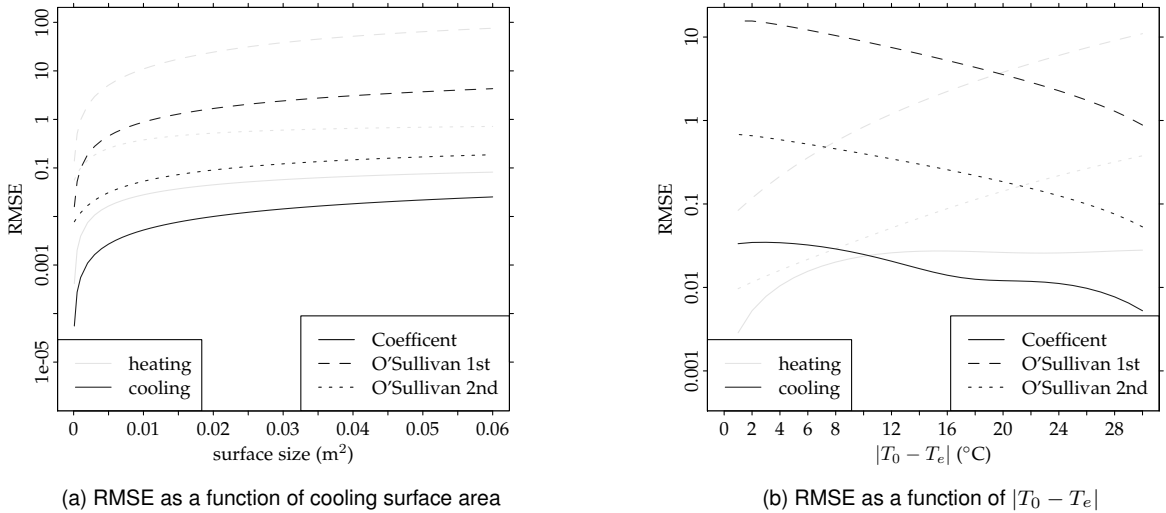

  \centering
  \subfloat[RMSE as a function of cooling surface area]{\input{approx-S-min.tex}%
  \label{fig:temp-delay:b}}
  \hfil
  \subfloat[RMSE as a function of $|T_0-T_e|$]{\input{approx-Te-min.tex}%
  \label{fig:temp-delay:d}}
  \caption{Root mean-square error {\smaller RMSE} between the accurate cooling law and the approximations. On the left (a) the surface area $S$ is variable, whereas the equilibrium temperature $T_e$ is variable in the right graph ($S=0.01$\,m$^2$) (b). The \emph{Coefficient Approximation} seems to outperform the other approximations. The \emph{Second-order O'Sullivan approximation} is performing acceptably as well for small values of $|T-T_a|$. {\color{black}Increasing the internal heat generation results in a decrease of the approximation error. We observe on average an overall ten-fold decrease between the maximum and minimum internal heat generation values.}}
  \label{fig:battle_of_the_approximations}
\end{figure*}

Figure~\ref{fig:battle_of_the_approximations} shows the \ac{RMSE} of the approximations for different surface areas, internal heat generation and equilibrium temperature settings.
From all graphs the coefficient approximation is clearly performing best.
Also, the second-order O'Sullivan approximation is considerably better than the first-order O'Sullivan approximation.
However, for very small surface area the errors in all approximations are acceptable.
Interestingly, the first-order O'Sullivan approximation does well for small surface areas, because the radiative part in the heat equation becomes negligible for smaller surface areas, and so the passive heat equations tends towards an exponential cooling law (see next section).
Consequently the first-order O'Sullivan approximation, being an exponential function, is able to approximate accurately the cooling law well for very small surface areas: $S < 0.005$\,m$^2$.

The errors for small internal heat generation seem to be systematically larger than the errors for the maximum internal heat generation case.
The same observation can be made for the heating and cooling processes.
The heating approximation seems to be more erroneous than the cooling process.

For variable equilibrium temperatures we see that for $|T_0-T_e|$ the error increases for the heating process and decreases for the cooling process.
In the derivation of the O'Sullivan approximations we have assumed that $T-T_a$ remains relatively small.
This implies that the larger $T$ departs from $T_a$ the more imprecise the approximation becomes.
For the cooling process $T_0$=55$^\circ$C and the equilibrium temperature $T_e$ was scaled between 25$^\circ$C and 55$^\circ$C.
Similarly, for the heating process $T_0$ was set to 25$^\circ$C and $T_e$ was scaled between 25$^\circ$C and 55$^\circ$C.
In both cases $T_a$ was fixed to 20$^\circ$C.
Thus as the cooling process approaches $T_a$ for increasing $|T_0-T_e|$, $T-T_a$ becomes smaller, and hence also the error between the O'Sullivan approximations and the exact cooling law.
The reverse observation is also valid for the heating process; \ac{RMSE} becomes larger for larger values of $T-T_a$.
The error properties in the case of the coefficient approximation is dependent on the fit of the second-order polynomial on the (quadratic) radiation function.

Overall, we do not advise to use the first-order O'Sullivan approximation, unless the surface area is really small, i.e., $\approx 0.005$\,m$^2$.
The second-order O'Sullivan approximation can be used but with caution.
The equilibrium temperature should not depart too much from the ambient temperature $T_a$; $T-T_a < 15^\circ$C seems acceptable.
We recommend, however, the use of the coefficient approximation, even though the solution isn't much elegant when the large polynomial coefficients are introduced.

\section{Comparison of the Passive and Active Cooling Laws Under Isothermal Conditions}
\label{sec:laws_differ}

Given the intrinsic complexity of the (inverse) function describing passive cooling compared to the rather straightforward exponential specification of other cooling modes, it is worth investigating in which cases dealing with it is necessary in practice.
We ran a large series of simulations to understand under what circumstances the passive and active cooling laws differ from each other.
{\color{black} The main difference between the active cooling (exponential-based) and the passive cooling law (see Section~\ref{sec:radcooling}) is the presence of the radiative heat transfer mode.}
Thus, if the radiative heat transfer is negligible compared to the convective heat transfer, the passive cooling law will approach an exponential cooling law.
We explore when such situations occur in concrete {\color{black}embedded system} use cases.

Let us recall that, for an isothermal body with internal heat generation, Equation~\ref{eq:solution-newton} governs active cooling and Equation~\ref{eq:solution-passive} governs passive cooling.
The internal heat generation $H(T)$ is a function of the temperature $T$.
We have shown that $H(T)$ is well described by an exponential equation~\cite{2014:devogeleer:samos}.
Even more, within the temperature range $25^\circ C < T < 55^\circ C$, the exponential can be approximated well with a linear or quadratic polynomial.
Yet, for the more extended temperature range $25^\circ C < T < 85^\circ C$, an exponential function is advised.

{\color{black}We compare the active and passive cooling of a system in the context of embedded devices, e.g., low-power {\small SoC}s or tablets subject to internal heating generation and cooling.
In order to do so, we assume a simplified system model: an isothermal volume with internal heat generation, cooled via convection and radiation.}
 \begin{table}
   \centering
   \small
   \caption{Variables used for the comparison of the active and passive cooling laws. The steady-state thermal behavior is analyzed. As a result,  \label{table:sim:settings}}
   \begin{tabular}{|c|c|c|} \hline
     \multicolumn{3}{|c|}{\sc Constants} \\\hline
     symbol & value & dim. \\\hline
     $\sigma$ & 5.670$\times$ 10$^{-8}$ & W/(m$^{2}$K$^{4}$)   \\
     $\epsilon$ & 0.94 & -  \\
     $T_a$ & 20 & $^\circ$C \\\hline
     \multicolumn{3}{|c|}{\sc Variables} \\\hline
     symbol & value & dim. \\\hline
     $S$ & $[0,6]\times$10$^{-3}$& m$^2$ \\
     $T$ & $[25,85]$  & $^\circ$C \\
     $h$ & (see Equation \ref{eq:hatTe-active}/\ref{eq:hatTe-passive}) & W/(m$^2$K) \\
     $\alpha_\text{min,max}$ & $\{0.396,4.030\}$ & W \\
     $\beta_\text{min,max}$  & $\{29.015,32.010\}$ & - \\
     $\gamma_\text{min,max}$ & $\{82.738,149.797\}$ & - \\\hline
   \end{tabular}
 \end{table} 
Table~\ref{table:sim:settings} shows the values used in our simulations.
The table lists the fixed variables: $\sigma$, $\epsilon$ and $T_a$.
We chose the emissivity of {\small PVC}\footnote{\color{black}Emissivity values of various packaging materials of embedded systems are often close to 0.95 to facilitate passive cooling.} for $\epsilon$ and fixed $T_a$ to be a representative room temperature.
The variables that may vary during the analysis are also listed.
We study the impact of the surface area $S$ over which the device cools via convection and radiation.
The minimum surface size was set to a square with a side of 1\,cm.
This is representative for a small \ac{SoC}; for example, the Samsung Exynos 5 \ac{SoC} has a side length of 1.6\,cm.
The maximum surface area was set to 0.06\,m$^2$, which is a representative area for a large tablet.
We analyze the behavior of the system within the temperature range $T\in[25,85]^\circ$C.
Throughout the analysis, we define the internal heat generation $H(T)$ to be an exponential function ($\alpha + e^{(T - \gamma) / \beta}$); the coefficients are shown in Table~\ref{table:sim:settings} as pairs.
The left values are for minimal internal heat generation, the right values for maximum internal heat generation.
The values for $\alpha$, $\beta$ and $\gamma$ were derived from power and temperature measurements on a \ac{SoC} sporting a {\small\sc CORTEX} A15~\cite{2014:devogeleer:samos}.
We measured the system's power consumption when the A15 is running at full capacity, i.e., at 1.6\,GHz, and when the A15 is running in low-power mode, i.e., at 800\,MHz.
{\color{black}
The heat capacity $C$ of the system is the product of its volume and its specific heat capacity and density.
In fact, in steady-state analysis the heat capacity does not affect the equilibrium temperature $T_e$.
Similarly, the performance metric $\Delta\tau$, used in Section~\ref{sec:temperature_differences}, is unaffected by $C$ as it merely scales equally the passive and active cooling processes in time.
This means that the data generated for the forthcoming Figures~\ref{fig:temp-delay} and \ref{fig:rcr} are independent of the actual composition of the body, being homogeneous or not, as long as the quasi-isothermal assumption holds.
Therefore, specific values for volume, specific heat capacity and density are not required in this section's analysis.
This analysis can thus apply to any system of arbitrary composition and size.

Also, note that the internal heat generation model used here only addresses the heat generated by a microprocessor.
In a more realistic setting, other components inside a computer system may also generate heat, e.g., radio interfaces, displays, \small{DC-DC} converters.
The heat generation model used here can be deemed as a lower bound on the actual internal heat generation of a practical embedded system.}

\subsection{Relative Heat Transfers}

First, we look at the ratio of the convective heat transfer coefficients of the passive and active cooling cases.
The temperature $T_0$ at $t$=0 is set to 25$^\circ$C.
Then we compute the respective convective heat transfer coefficients as per Equation~\ref{eq:hatTe-passive} and Equation~\ref{eq:hatTe-active} based on a series of equilibrium temperatures $T_e$.
The ratio $r_\mathrm{cr}$ of the convective heat transfer coefficients is given by
\begin{equation*}
r_\mathrm{cr} = \frac{h_\mathrm{pc}}{h_\mathrm{ac}} = \frac{\epsilon \sigma S (T_a^4 - T_e^4) + H(T_e)}{H(T_e)}.
\end{equation*}
$r_\mathrm{cr}$ shows how much the active and passive cooling laws will resemble.
If $r_\mathrm{cr}=1$, there is no difference between the two cooling cases.
The more $r_\mathrm{cr}$ tends to zero, the more the two cooling laws will deviate in behavior.

\begin{figure*}[!t]
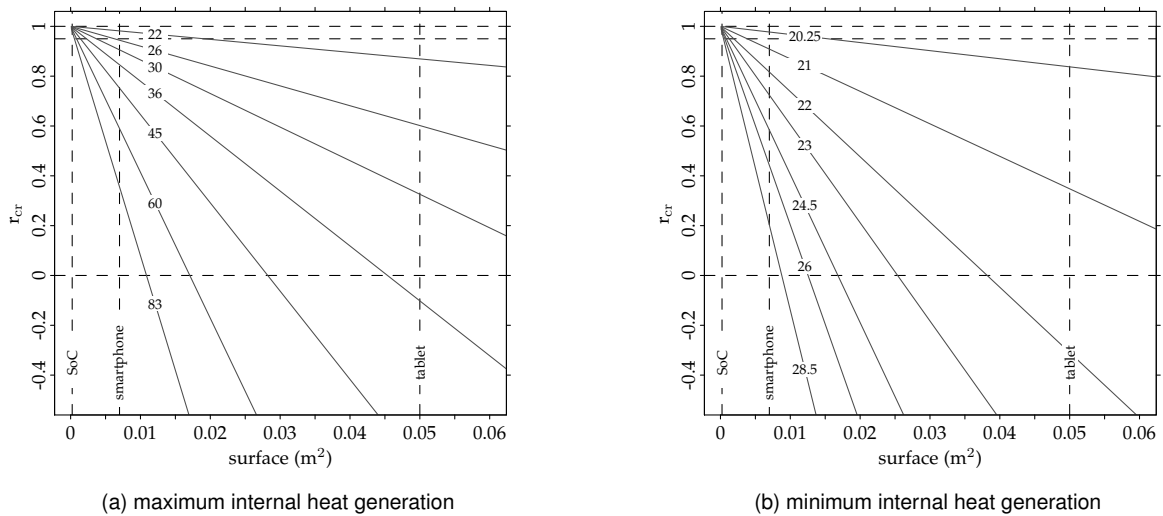

  \centering
  \subfloat[maximum internal heat generation]{\input{hpchac-large.tex}%
  \label{fig:gull}}
  \hfil
  \subfloat[minimum internal heat generation]{\input{hpchac-small.tex}%
  \label{fig:tiger}}
  \caption{{\color{black}Ratio between the convective heat transfer coefficients of active and passive cooling, at a given equilibrium temperatures $T_e$ [$^\circ$C] (curve labels) and ambient temperature $T_a$ of 20$^\circ$C.} In Figure (a) the internal heat generation is set to a maximum, while in (b) it is set to a minimum, following Table~\ref{table:sim:settings}. The vertical dashed lines represent typical surfaces of a {\smaller SoC} ($\approx 2.5$\,cm$^2$), a smartphone ($\approx 70$\,cm$^2$) and a tablet ($\approx 5$\,dm$^2$). A horizontal reference line is drawn at $r_\mathrm{cr} = 0.95$.}
  \label{fig:rcr}
\end{figure*}

Figure~\ref{fig:rcr} shows the ratio of the convective heat transfer coefficient of the passive and active cooling cases.
Given that $r_\mathrm{cr}$ stays well above 0.95, it is observed that, for a small system, similar to an \acp{SoC} (left most vertical dashed line), the difference between active and passive cooling will be very small for all equilibrium temperatures ranging between 20$^\circ$C to 85$^\circ$C.
For a moderate surface area, e.g., the size of an average smartphone (middle vertical dashed line), the radiative cooling starts to become more prominent already for temperatures close to the ambient temperature $T_a$.
For equilibrium temperatures more than about 5$^\circ$C above $T_a$, signs of deviating behavior will become clearly visible.
Large surface areas and equilibrium temperatures close to $T_a$ will yield a $r_\mathrm{cr}$ that is smaller than 0.95.
This implies that the radiative cooling for large surfaces has definitely to be taken into account.
As a general rule of thumb, we can say that the larger the equilibrium temperature and the cooling surface, the more behavioral differences between passive and active coolings will occur.
So how large are the differences temperature-wise in particular?

\subsection{Temperature Differences}
\label{sec:temperature_differences}

When looking at the temperature differences between the passive and active cooling laws at specific points in time, we must differentiate between the cooling and heating processes.
Convective heat transfer is proportional to the difference of the {\color{black}system's} temperature and the ambient temperature, and is therefore \emph{independent} on the absolute temperature of the {\color{black}system} and environment.
This results in a symmetry between the heating and the cooling processes for convective heat transfer.
The radiative heat transfer, on the other hand, is \emph{dependent} on the absolute values of the body and the environment.
This is illustrated as follows for the convective and radiative heat transfers respectively:
\begin{eqnarray}
| h S (T - (T-x)) | & = & | h S( T - (T+x)) | \nonumber\\
| \epsilon \sigma S (T^4 - (T-x)^4) | & \neq & | \epsilon \sigma S (T^4 - (T+x)^4) |\label{eq:convective-inequality}
\end{eqnarray}
As a consequence, due to the last inequality, the radiative heat transfer process will not be symmetric for the cooling and  heating processes.
Moreover, when radiative heat transfer is combined with convective heat transfer, the symmetry property of the heating and cooling processes will not hold either.

\begin{figure*}[!t]
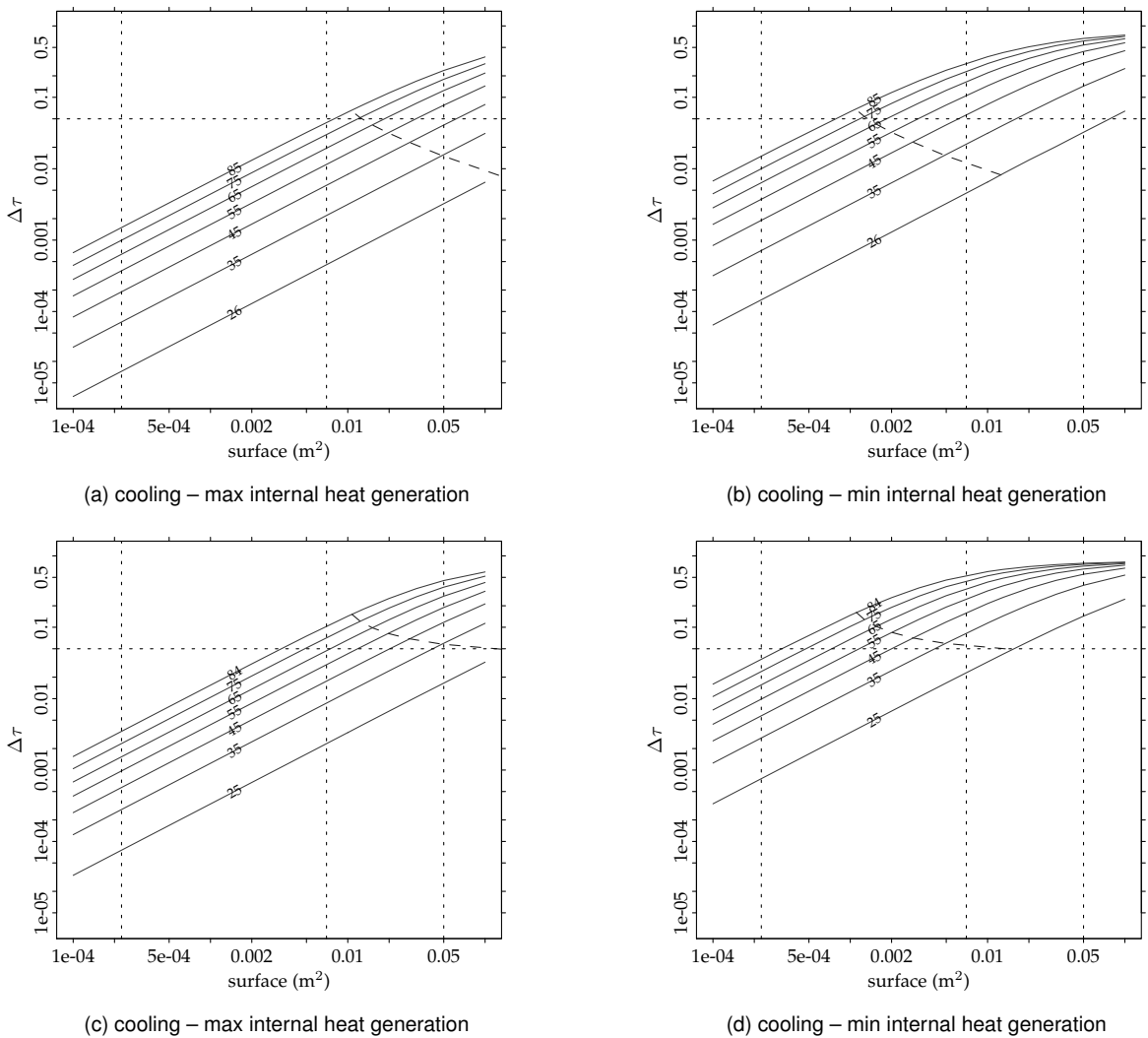

  \centering
  \subfloat[cooling -- max internal heat generation]{\input{dtdelay-ilarge-heat.tex}%
  \label{fig:temp-delay:a}}
  \hfil
  \subfloat[cooling -- min internal heat generation]{\input{dtdelay-ismall-heat.tex}%
  \label{fig:temp-delay:b}}
  \hfil
  \subfloat[cooling -- max internal heat generation]{\input{dtdelay-ilarge-cool.tex}%
  \label{fig:temp-delay:c}}
  \hfil
  \subfloat[cooling -- min internal heat generation]{\input{dtdelay-ismall-cool.tex}%
  \label{fig:temp-delay:d}}
  \caption{Relative time lag $\Delta\tau$ (Equation~\ref{eq:deltatau}) for the internal heat generation set to the maximum (a,c), and set to the minimum (b,d). The  curves are generated for different equilibrium temperatures (see curve labels in $^\circ$C). On the top row, the heating process is depicted (a,b), with the cooling process on the bottom row (c,d). The three vertical dotted lines represent typical surfaces for a {\smaller SoC} ($\approx 2.5$\,cm$^2$), a smartphone ($\approx 70$\,cm$^2$) and a tablet ($\approx 5$\,dm$^2$). Data points on the right of the blue dashed lines have negative convective heat transfer coefficients.}
  \label{fig:temp-delay}
\end{figure*}

Let us define the temperature lag $\Delta T$ between two actively and passively cooled identical {\color{black}systems}, measured at the moment when the passively cooled {\color{black}system} reaches a reference temperature $T_\mathrm{pc}$.
The reference temperature $T_\mathrm{pc}$ is henceforth defined as $T_\mathrm{pc}=0.85(T_e-T_0)+T_0$, i.e., when the {\color{black}system}'s temperature has reached 85\% of its equilibrium temperature, starting from $T_0$.
It is also assumed that both the passively and actively cooled {\color{black}systems} have the same internal heat generation process and initial condition $T_0$ at $t=0$.
Figure~\ref{fig:temp-delay} shows the relative temperature lag $\Delta \tau$, which is defined as the absolute temperature lag $\Delta T$ divided by the temperature difference at $t=0$ and at equilibrium $|T_e - T_0|$:
\begin{equation}
 \Delta \tau = \frac{\Delta T}{|T_e-T_0|} = \frac{T_\mathrm{pc} - T_\mathrm{ac}}{|T_e-T_0|}.\label{eq:deltatau}
\end{equation}
{\color{black}Even though the definition of $\Delta \tau$ here is time-independent, it does tell us something about the difference in transient behavior between passive and active cooling processes.}

The relative temperature lag $\Delta\tau$ is depicted in Figure~\ref{fig:temp-delay} for both a large and a small internal heat generation, as defined before, and for the heating and cooling processes separately.
{\color{black}The smaller $\Delta \tau$, the better.}
A reference line is drawn for $\Delta \tau = 5\%$.
Data points on the right of the dashed blue line show configurations with one or more negative convective heat transfer coefficients.
This implies that in these cases additional heat needs to be added to attain the given equilibrium temperature.
These data points are however, not of concern in our work.

For the case of large internal heat generation, the relative temperature lag $\Delta \tau$ for small surfaces stays below 0.5\%, meaning that the presence of radiative heating will be quasi unnoticeable here.
$\Delta\tau$ stays around 5\% in the case of small internal heat generation, which may be difficult to spot.
{\color{black}Contemporary embedded system temperature sensors, e.g., on-die microprocessor sensors, report frequently temperature values in steps of 1$^\circ$C.}
Given this quantization noise, a relative temperature lag of 5\% could be hard to identify when $|T_e-T_0|>20^\circ$C.
So for small system temperature variations, it is again unlikely that a contemporary temperature sensor is able to distinguish between active and passive cooling.
For a smartphone-size cooling surface, the relative temperature lag varies significantly depending on the situation.
For a large internal heat generation and heating, there is less than 5\% difference between passive and active cooling.
For the other cases, however, the discrepancy between the passive and active cooling can run up from nil to as high as 10\%, depending on the equilibrium temperature.
$\Delta\tau = 10\%$ is already noticeable at $|T_e-T_0|>10^\circ$C in the presence of 1$^\circ$C quantization noise.
{\color{black}The data for the tablet-sized cooling surfaces shows that the temperature difference between active and passive cooling can become as high as 50\%.
This implies that for the larger embedded systems radiative cooling should definitely be considered when designing a realistic thermal profile of the system.}

Generally speaking, we notice that the relative temperature lag $\Delta\tau$ for heating cases is smaller than for the cooling cases.
This can be explained via the inequality of Equation~\ref{eq:convective-inequality}.
The radiative heat transfer coefficient will have greater weight when the {\color{black}system}'s temperature is larger than the equilibrium temperature than when the temperature is below the equilibrium, hence inflating the discrepancy between active and passive cooling.
Also, the amount of internal heat generation affects the relative temperature lag.
It appears that the larger the internal heat generation, the smaller $\Delta\tau$ becomes.
Indeed, given the differential representation of the cooling law in Equation~\ref{eq:start}, for a fixed equilibrium temperature, we see that the convective cooling part can outweigh the radiative the larger the internal heat generation becomes.
Thus the larger the internal heat generation, the less sensitive the {\color{black}system} becomes to changes in the radiative or convective cooling, and the more active and passive cooling will resemble.

\section{Conclusion}
\label{sec:conclusion}

{\color{black}We have introduced a new, more accurate cooling law for passively cooled embedded system-like devices subject to radiation, (natural) convection, and internal heat generation.}
The passive cooling law is analytically more complex than the commonly accepted exponential cooling law (which is technically sound for forcibly cooled {\color{black}systems}).
Unfortunately, the accurate solution for the passively cooled {\color{black}system} is a function of temperature: $t(T)$.
{\color{black}Either numerical approaches can be used to compute the exact inverse: $T(t)$, or one of our presented approximations can generate a good enough approximation to the cooling law.
The validation of the passive cooling law's accurate solution via \ac{CFD} and electrical-equivalence simulations demonstrated the cooling law's practical adequacy.}

Via analytical simulations, we showed that the difference between active and passive cooling depends on three factors: 1) the surface area of the object, 2) the internal heat generation, and 3) the equilibrium temperature.
For large objects, we showed that the difference between active and passive cooling can be significant.
For medium-sized ones, depending on the magnitude of the internal heat generation and equilibrium temperature, the discrepancy between active and passive cooling could tentatively go unnoticed.
For small surfaces, e.g., \acp{SoC}, an exponential cooling law is shown to be an appropriate approximation.
We also highlighted that the quantization noise of temperature sensors may conceal temporal information between active and passive cooling.
As the cooling law for passively cooled devices is quite elaborate to work with and the possible uses of a scientifically sound cooling law by \acp{TMU} are limited by the lack of accurate temperature sensors, we can state that, for systems minimizing overhead, assuming an exponential cooling law will likely not induce large perceptual deviations from reality.
As the cooling law for passively cooled devices is quite elaborate to work with and the possible uses of a scientifically sound cooling law by \acp{TMU} are limited by the lack of accurate temperature sensors, we can state that, for systems minimizing overhead, assuming an exponential cooling law will likely not induce large perceptual deviations from reality.

In this work we considered the cooling of an isothermal object.
In practical situations this assumption doesn't always hold.
To obtain a more realistic model we need to consider internal conduction, and hence also thermal hotspots.
The impact of these considerations on our heat model is part of our future work.
{\color{black}Moreover, embedded systems consist of multiple subsystems, e.g., a microprocessor, \acp{PCB}, and are covered by other objects, such as an {\small LCD} display, radio interface and others.}
The presence of these objects also interacts with the passive cooling of the entire computer system.
Most likely numerical methods will have to be deployed to gain a more {\color{black}detailed} understanding under such conditions.

\ifCLASSOPTIONcompsoc
  \section*{Acknowledgments}
\else
  \section*{Acknowledgment}
\fi

We would like to thank the staff at the Department of Bioengineering at Ghent University for the support with the simulation aspects of this work.

\appendices

\section{Applicability of the Passive Heat Equation}
\label{sec:Applicability}


Previously we assumed that the internal heat generation $H(T)$ was a linear function, i.e., polynomial of the first-order with coefficients elements of $\mathbb{R}^+$.
Given that the radiation absorbed or emitted by a body is described by a 4th-order polynomial, we discuss the implications of an arbitrary $H(T)$ up to the 3th-order.
We will show via logical reasoning that the analytic solution in the paper holds for $H(T)$ up to the 3th order under certain conditions.

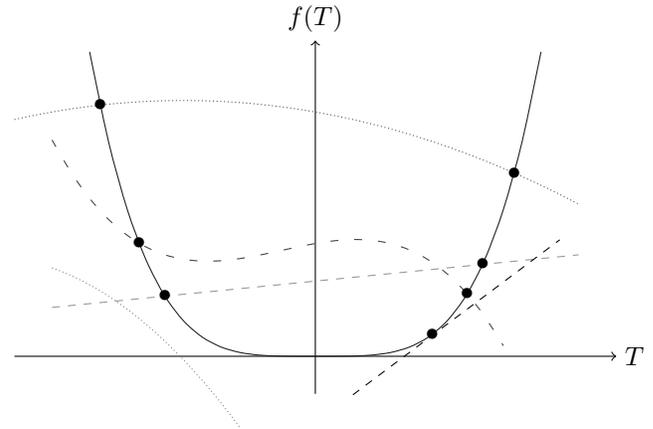
\begin{figure}
	\begin{center}

\begin{tikzpicture}[scale=1.0]
  \draw[->] (-4,0) -- (4,0) node[right] {$T$};
  \draw[->] (0,-0.5) -- (0,4.2) node[above] {$f(T)$};
  \draw[scale=0.05,domain=-3:3,smooth,variable=\x,blue] plot ({\x/0.05},{(\x)^4});
  \draw[scale=0.05,domain=-3.5:3.5,smooth,variable=\x,green,dashed] plot ({\x/0.05},{2*(\x)+20});
  \draw[scale=0.05,domain=0.5:3.25,smooth,variable=\x,black,dashed] plot ({\x/0.05},{15*(\x)-17.75});
  \draw[scale=0.05,domain=-4:3.5,smooth,variable=\x,red,densely dotted] plot ({\x/0.05},{-(\x)^2-3.5*(\x)+65});
  \draw[scale=0.05,domain=-3.5:-1,smooth,variable=\x,gray,densely dotted] plot ({\x/0.05},{-4*(\x)^2-35*(\x)-50});
  \draw[scale=0.05,domain=-3.5:2.5,smooth,variable=\x,purple,loosely dashed] plot ({\x/0.05},{-1.5*(\x)^3-2*(\x)^2+3.5*(\x)+30});
  
  \foreach \Point in {(2.223603,24.44719*0.05), (-2,16*0.05), (1.552457,5.808692*0.05), (2.018136,16.58829*0.05), (-2.342428,30.10683*0.05), (-2.859212,66.83215*0.05), (2.642607,48.76748*0.05)}{
    \node at \Point {\textbullet};
  }
  
\end{tikzpicture}
	
		\caption{Visualization of Equation~\ref{eq:HTeq} for several variations of the right-hand side polynomial ($K(T)$).
			  Polynomials: 1st order (dashed), 2nd order (dotted), 3rd order (loosely dashed), and $\delta T^4$ (solid).
			  The black bullets represents the intersections of each polynomial with $\delta T^4$.}
		\label{fig:HT}
	\end{center}
\end{figure}

Let us define a body that is radiating energy at a rate $-\delta$, and subject to other heat transfer mechanisms described by a polynomial $K(T)$, e.g., internal heat generation.
Let $K(T)$ be a polynomial of an order not larger than three.
Then the thermal energy storage rate into the body is equal to:
\begin{equation}
  C \frac{dT}{dt} = -\delta T^4 + K(T)  =  -\delta T^4 + (\kappa_3 T^3 + \kappa_2 T^2 + \kappa_1 T + \kappa_0),\label{eq:KT}
\end{equation}
where we define $\delta\in\mathbb{R}^+_0$, $\kappa_{0,1,2,3}\in\mathbb{R}$, and $C$ is the thermal capacity of the system.
$\delta$ must be positive as $-\delta T^4$ represents the heat emitted by the body via radiation.
$\kappa_{0,1,2,3}$ are the constants of a polynomial describing the function $K(T)$.
To solve the differential in Equation~\ref{eq:KT} the roots need to be found.
In particular, we have solved the differential equation for a 4th-order polynomial assuming two real and two complex conjugate roots.
To find the roots of Equation~\ref{eq:KT} we evaluate it at the equilibrium temperature $T=T_e$, then $dT/dt=0$:
\begin{equation}
  \delta T^4 = \kappa_3 T^3 + \kappa_2 T^2 + \kappa_1 T + \kappa_0.\label{eq:HTeq}
\end{equation}
This equality is visualized in Figure~\ref{fig:HT}.
There the solid blue curve represents the contribution on the left-hand side and the other dashed lines are possible examples of the polynomial in the right-hand side.
It can be seen that it is easy to construct polynomials that have one or two intersections with $\delta T^4$.
Also curves can be constructed that intersect the $\delta T^4$ only in one point (for example the dashed black line in Figure~\ref{fig:HT}); such points are counted as two roots.
The dashed gray line is an example of a polynomial without any intersection with $\delta T^4$.
Only those polynomials with one or two intersections with $\delta T^4$ have physical meaning in the context discussed in this paper.
One or two intersections with $\delta T^4$ produce two real roots and two complex conjugate roots.
No intersections with $\delta T^4$ would imply that there exists no equilibrium temperature, i.e., the system is not thermally stable.

\section{Solving the Passive Heat Equation}
\label{sec:SolutionPassiveHeating}

The differential formulation of a passively cooled object with linear internal heat generation can be described as follows, as per Equation~\ref{eq:diff-passive}:
\begin{equation*}
\frac{\diff T}{\diff t} = \frac{1}{C} (- \epsilon \sigma S T^4  +  ( \eta_1 - hS) T + ( \eta_0 + S( h T_a + \epsilon \sigma  T^4_a ))).
\end{equation*}
The right-hand side is a fourth-order polynomial and the equality can be rephrased as:
\begin{equation}
\frac{\diff T}{\diff t} = -\kappa_4 T^4 + \kappa_3 T^3 + \kappa_2 T^2 + \kappa_1 T + \kappa_0,\label{eq:pokjmnsd}
\end{equation}
where the constants $\kappa_4\in\mathbb{R}^+_0$ and $\kappa_{\{0,1,2,3\}}\in\mathbb{R}^+$.
Rearranging this equation yields
\begin{eqnarray}
\int \frac{1}{ T^4 - \frac{\kappa_3}{\kappa_4} T^3 - \frac{\kappa_2}{\kappa_4} T^2 - \frac{\kappa_1}{\kappa_4} T - \frac{\kappa_0}{\kappa_4}} \diff T & = & -\kappa_4 \int \diff t  \label{eq:kappa-sides}.
\end{eqnarray}
The integration of the fraction on the left-hand side can be achieved via partial fractions decomposition:
\begin{equation}
\int\frac{1}{(T-\omega_1)(T-\omega_2)(T-\omega_3)(T-\omega_4)} \diff T \label{eq:init}
\end{equation}
The roots $\omega_*$ of the 4th order polynomial in the denominator can be obtained via Ferrari's theorem, and other approximate methods such as Netwon's and the secant.
Given that there exist a maximum of one or two real unique values for $T$ that satisfy
\begin{equation*}
\kappa_4 T^4 = \sum_{i=0}^3 \kappa_i T^i,
\end{equation*}
we can state that two roots are real, say $\omega_{\{1,2\}}$; the other two roots are complex conjugates\footnote{Appendix~\ref{sec:Applicability} shows that for our applications this is the case.}.
This means that $\Re(\omega_3)=\Re(\omega_4)$ and $\Im(\omega_3)=-\Im(\omega_4)$, which simplifies a few things.
As the initial differential equation is real, we are looking for a real solution too; thus the imaginary part must equate to zero.
This is however automatically taken care of as the product of the two complex roots yield a real sum:
\begin{equation*}
\frac{1}{(T-\omega_3)(T-\omega_4)} = \frac{1}{(T-\Re(\omega_3))^2+\Im(\omega_3)^2}.
\end{equation*}
Whence, Equation \ref{eq:init} becomes
\begin{equation}
\int\frac{A}{(T-\omega_1)}+\frac{B}{(T-\omega_2)}+\frac{C T + D}{(T-\Re(\omega_3))^2+\Im(\omega_3)^2} ~\diff T. \label{eq:split}
\end{equation}
Henceforth we define $\alpha=\Re(\omega_3)$ and $\beta=\Im(\omega_3)$.
The values for $A$, $B$, and $D$ are found by equating Equation~\ref{eq:init} and Equation~\ref{eq:split}, which can be expressed as a system of equations:
\begin{eqnarray*}
 \begin{cases}
  0 & =~ A+B+C \\
  0  & =~ D-\oa (B+C)-\ob (A+C)-2\rr(A+B) \\
  0  & =~ \rr^2 (A+B) +\beta^2(A+B) +2\rr (\ob A + \oa B) \\
     & \quad \quad \quad -(\oa+\ob) D +\oa\ob C \\
  1  & =~ -\rr^2 ( \ob A + \oa B) - \beta^2 (\ob A + \oa B) +\oa\ob D
  \end{cases}
\end{eqnarray*}
and can be solved via Gaussian elimination.
So we obtain the expressions for $A$, $B$, $C$ and $D$:
\begin{subequations}
\label{eq:constants}
\begin{empheq}{align}
   A  & =~ \frac{1}{ (\oa-\ob)((\rr^2+\beta^2)-\oa (2\rr-\oa))} \\
   B  & =~ -A \frac{\rr^2 + \beta^2 - \oa (2\rr-\oa) }{\rr^2 + \beta^2  -  \ob (2\rr- \ob) }   \\
   C  & =~ -(A+B) \\
   D  & =~ A (2\rr-\oa) + B (2\rr- \ob)
\end{empheq}
\end{subequations}

Continuing with Equation~\ref{eq:split}, this yields:
\begin{equation*}
A\ln|T-\oa | + B\ln|T-\ob | + \int\frac{CT+D}{(T-\alpha)^2+\beta^2} \diff T + c_0,
\end{equation*}
where $c_o$ is an integration constant.
The last term on the right-hand side may be integrated via substitution, where $u=(T-\rr)^2$, yielding $du = 2(T-\rr)\diff T$, and also $v=\frac{T-\alpha}{\beta}$, giving $dv = \frac{1}{\beta}\diff T$:
\begin{eqnarray*}
& & \int\frac{CT+D}{(T-\alpha)^2+\beta^2} \diff T  \\
& & = \int\frac{C(T-\alpha)}{(T-\alpha)^2+\beta^2} \diff T +\int\frac{\alpha C+D}{(T-\alpha)^2+\beta^2} \diff T \\
					&  &  = \frac{C}{2} \ln| (T-\rr)^2 + \beta^2| +  \frac{\alpha C+D}{\beta}\arctan\left( \frac{T-\rr}{\beta} \right) + c_1.
\end{eqnarray*}
where $c_1$ is an integration constant. Then the solution to Equation~\ref{eq:split} is as follows
\begin{eqnarray}
& & A\ln|T-\oa | + B\ln|T-\ob | + \frac{C}{2} \ln| (T-\rr)^2+ \beta^2| \nonumber\\
&  & \quad\quad   +  \frac{\alpha C+D}{\beta}\arctan\left( \frac{T-\rr}{\beta} \right) + c_1,
\end{eqnarray}
where $A$, $B$, $C$ and $D$ are given in Equations~\ref{eq:constants}, and $\omega_*$ are the real roots of the polynomial in the denominator on the left-hand side, $\alpha=\Re(\omega_3)$, $\beta=\Im(\omega_3)$, and $c_1$ is a (new) integration constant satisfying the initial conditions.

Now we can complete Equation~\ref{eq:kappa-sides}:
\begin{eqnarray*}
 t & = & -\frac{1}{\kappa_4}\bigg(A\ln|T-\oa |  + \frac{C}{2} \ln| (T-\rr)^2+ \beta^2|  + c_o  \nonumber\\
 &  & \quad\quad  + B\ln|T-\ob | +  \frac{\alpha C+D}{\beta}\arctan\left( \frac{T-\rr}{\beta} \right)\bigg).
\end{eqnarray*}

\section{Derivations of Approximations for $f(T)=t$}
\label{sec:approx_derivation}

The exact passive cooling law as presented is of the form $f(T)=t$.
For practical reasons we desire a formulation of the form $f(t)=T$.
Unfortunately inverting the exact passive heat equation is challenging.
We develop three approximations to the exact passive cooling law which are more easily invertible.

\subsection{Quadratic Approximation}
Stefan-Boltzmann's law of radiation states that the energy emitted by radiation is proportional to  $T^4$ (Equation \ref{eq:Stefan-Boltzmann-law}).
Because of this term the polynomial of Equation~\ref{eq:diff-passive} is of the fourth-order.
More specifically, it are the two imaginary roots of the fourth order polynomial that introduce the \emph{arctan} in Equation~\ref{eq:solution-passive}.
If we were to approximate $T^4$ with a second-order polynomial and assert real roots, then we could get rid of the dependency of the \emph{arctan}, and isolating $T$ would be more straightforward.
The quadratic approximation
\begin{eqnarray}
	T^4 & = & q_0 + q_1 T +  q_2 T^2 \nonumber \\
		& = & 29700057265  - 251483462\,T +  598262\,T^2\label{eq:temperature-approximation}
\end{eqnarray}
introduces an error between -0.041\% and 0.072\% for $20^\circ$C $< T < 65^\circ$C, which is very acceptable.
Then the quadratic approximation to Equation~\ref{eq:pokjmnsd} would be equal to solving
\begin{equation}
	\frac{dT}{dt} = \kappa_2 T^2 + \kappa_1 T + \kappa_0.\label{eq:approx:quadratic:begin}
\end{equation}
The solution to this equation, assuming two real roots ($\omega=(-\kappa_1 \pm \sqrt{\kappa_1^2-4\kappa_2\kappa_0}/(2\kappa_2))$) and that $\kappa_2 < 0$:
\begin{equation}
	t = -\frac{1}{\kappa_2}\left(A\ln|T-\oa | + B\ln|T-\ob | + c_o \right),\label{eq:thesolutionapprox}
\end{equation}
where $A = 1/(\omega_2-\omega_1)$ and $B = -A$.
Now we can isolate $T$ as follows:
\begin{eqnarray*}
	t + \frac{c_o}{\kappa_2}& = & -\frac{A}{\kappa_2}\left(\ln|T-\oa | - \ln|T-\ob | \right)  \\
	-\frac{\kappa_2 t + c_o}{A} & = & \ln\left( \frac{|T-\oa |}{|T-\ob |} \right) \\
	c_oe^{-\frac{\kappa_2}{A}t } & = & \frac{|T-\oa |}{|T-\ob |}.
\end{eqnarray*}
Let's define $\omega_1$ and $\omega_2$ such that $\omega_1 < \omega_2$.
As we are operating in the temperature range $0^\circ \text{C} < T < 100^\circ\text{C}$ and given the shape of the quadratic approximation, $T$ will always be larger than $\omega_1$.
Hence we can assume that $T-\oa > 0$.
The absolute value of $T-\omega_2$ forces us to distinguish two cases, i.e. where $T > \omega_2$ and the case for $T < \omega_2$.
Bear in mind that $\omega_2$ is also the equilibrium temperature $T_e$ of the system.
This corresponds either to the heating or the cooling process, respectively.
For $T > \omega_2$ we have
\begin{eqnarray}
	T-\oa & = & (T-\ob) c_o e^{-\frac{\kappa_2}{A}t} \nonumber\\
	T & = & \frac{\oa-\ob c_o e^{-\frac{\kappa_2}{A} t}}{1-c_oe^{-\frac{\kappa_2}{A}t }} \nonumber
\end{eqnarray}
and accordingly for $T < \omega_2$, or the heating process, we get:
\begin{equation}
	T = \frac{\oa+\ob c_o e^{-\frac{\kappa_2}{A} t}}{1+c_oe^{-\frac{\kappa_2}{A}t }},\label{eq:approx:quad}
\end{equation}
where $c_o$ is an integration constant to meet the initial condition $f(0)=T_0$, and given by \begin{equation*}
c_o = \frac{|T_0-\omega_1|}{|T_0-\omega_2|}.
\end{equation*}
The roots $\omega_*$ are are easily found as follows:
\begin{equation*}
 \omega_1 = \frac{-\kappa_1 + \sqrt{\kappa_1^2-4 \kappa_2 \kappa_0}}{2\kappa_2} \quad\text{and}\quad \omega_2 = \frac{-\kappa_1 - \sqrt{\kappa_1^2-4 \kappa_2 \kappa_0}}{2\kappa_2}.
\end{equation*}
The equilibrium temperature $T_e$ is defined by the positive root $\omega_2$.

In the above derivation, we have fixed the coefficients $q_*$ in Equation~\ref{eq:temperature-approximation}.
These values were chosen to fit best in a certain temperature range.
To be more universally applicable, however, the coefficients could be generated dynamically such that they are optimally tailored to the temperature range of concern.

\subsection{First Order O'Sullivan Approximation}

O'Sullivan~\cite{Sullivan:newton} presented an approximation for a cooling law including convection and radiation, but without the presence of internal heat generation.
We extend his approximation with internal heat generation.
We will use an alternative formulation of the internal heat generation such that we can more easily apply our variable substitution later on: $H(T)=\eta_1 T + \eta_0 = \eta_1 (T-T_a) + \eta_1 T_a + \eta_0$.
The initial definition of the passive heat Equation~\ref{eq:diff-passive} then becomes:
\begin{eqnarray*}
 -C \frac{\dd T}{\dd t} & = & \epsilon\sigma S (T^4-T_a^4) + (hS-\eta_1)(T-T_a) \\
 &  &  \quad - ( \eta_1 T_a + \eta_0).
\end{eqnarray*}
Let's introduce the variable $\theta = T-T_a$:
\begin{equation*}
 -C \frac{\dd \theta}{\dd t} = \epsilon\sigma S ((\theta + T_a)^4-T_a^4) + (hS-\eta_1)\theta - ( \eta_1 T_a + \eta_0).
\end{equation*}
Now, we can apply binomial expansion to $(\theta - T_a)^4$, whence:
\begin{eqnarray}
 -C \frac{\dd \theta}{\dd t} & = & \epsilon\sigma S ((\theta^4 + 4T_a\theta^3 + 6T_a^2\theta^2 + 4 T_a^3 \theta \nonumber\\
 & & \quad + T_a^4)-T_a^4) + (hS-\eta_1)\theta - ( \eta_1 T_a + \eta_0) \nonumber\\
 & = & k\theta^4 + l\theta^3 + m\theta^2 + n\theta + p, \label{eq:approx:osilluvian}
\end{eqnarray}
where the coefficients for surfaces around 1\,dm$^2$ are as follows:
\begin{eqnarray*}
k & = & \epsilon\sigma S \quad(\sim 10^{-10})\\
l & = & 4\epsilon\sigma ST_a \quad(\sim 10^{-7})\\
m & = & 6\epsilon\sigma S T_a^2 \quad(\sim 10^{-5})\\
n & = & (hS-\eta_1+4 \epsilon\sigma S T_a^3) \quad(\sim 0.01)\\
p & = & -( \eta_1 T_a + \eta_0) \quad(\sim 1).
\end{eqnarray*}
Now, if $(T-T_a)$ is not too large the series on the right-hand side of Equation~\ref{eq:approx:osilluvian} converges reasonably fast~\cite{Sullivan:newton}.
Depending on the accuracy desired, the higher-order coefficients may be dropped.
Let's see how well a first-order and a second-order approximation behaves.
As expected, the first-order approximation yields also an exponential law:
\begin{equation*}
-C \frac{\dd \theta}{\dd t} =  n\theta + p \quad \Rightarrow \quad \theta = c_oe^{-\frac{n}{C}t} - \frac{p}{n},
\end{equation*}
where $c_o$ is an integration constant such that $\theta(t=0)=T_0-T_a$:
\begin{equation*}
 c_o = \theta_0+\frac{p}{n} = (T_0-T_a) + \frac{p}{n}.
\end{equation*}
And so the first-order O'Sullivan solution is:
\begin{equation}
T = \left(T_0-T_a + \frac{p}{n}\right) e^{-\frac{n}{C}t} - \frac{p}{n} + T_a.\label{eq:approx:osullivan:first}
\end{equation}



\subsection{Second-Order O'Sullivan Approximation}

The second-order O'Sullivan approximation is a bit more complex compared to the first-order O'Sullivan approximation.
Moreover, the derivation looks also significantly different from the original derivation of O'Sullivan~\cite{Sullivan:newton}, given the presence of the constant term $p$ in Equation~\ref{eq:approx:osilluvian}.
The second-order O'Sullivan approximation is similar to the coefficient approximation in the sense that solving
\begin{equation}
 -C\frac{\dd \theta}{\dd t} = m\theta^2 + n\theta + p\label{eq:approx:osullivan:begin}
\end{equation}
is similar to solving Equation~\ref{eq:approx:quadratic:begin}.
Thus the solution for the second-order O'Sullivan approximation will be the same as for the quadratic approximation, except for the constants definition.
We can thus state that the second-order O'Sullivan approximation is given by:
\begin{equation}
 T = \frac{\omega_1 \pm \omega_2c_oe^{-\frac{m}{AC}t}}{1\pm c_oe^{-\frac{m}{AC}t}}+T_a,\label{eq:approx:osullivan:second:solution}
\end{equation}
where ''$\pm$'' becomes ''$+$'' for $T_e > T_0$, and ''$-$'' for $T_e < T_0$.
$\omega_*$ is given by:
\begin{equation*}
  \omega_1 = \frac{-\sqrt{n^2-4pm}-n }{2m} \quad \text{and} \quad \omega_2 = \frac{\sqrt{n^2-4pm}-n}{2m}.
\end{equation*}
The constant $A$ and $c_o$, such that $\theta(0)=\theta_0$, are defined as:
\begin{equation*}
  A = -\frac{1}{\omega_2-\omega_1}\quad \text{and} \quad c_o = \frac{|\theta_0-\omega_1|}{|\theta_0-\omega_2|},
\end{equation*}
where $\theta_0=T_0-T_a$.
The equilibrium temperature $T_e$ is defined by $\omega_2+T_a$.

\ifCLASSOPTIONcaptionsoff
  \newpage
\fi



\bibliographystyle{IEEEtran}
%
\bibliography{library}

\end{document}